\documentclass[12pt]{article}
\usepackage{times}
\usepackage{geometry}
\usepackage[utf8]{inputenc}
\usepackage{enumitem,amssymb}
\usepackage{ragged2e}
\usepackage{graphicx,xspace,lscape,floatflt,wasysym}
\usepackage{natbib}
\newlist{thematic}{itemize}{8}
\setlist[thematic]{label=$\square$}
\usepackage{pifont}

\begin{document}
\RaggedRight
\huge
Astro2020 Science White Paper \linebreak

Probing the Structure of Interstellar Dust from Micron to Kpc Scales with X-ray Imaging\linebreak
\normalsize

\noindent \textbf{Thematic Areas:} \hspace*{60pt} $\square$ Planetary Systems \hspace*{10pt} $\XBox$ Star and Planet Formation \hspace*{20pt}\linebreak
$\square$ Formation and Evolution of Compact Objects \hspace*{31pt} $\square$ Cosmology and Fundamental Physics \linebreak
  $\square$  Stars and Stellar Evolution \hspace*{1pt} $\square$ Resolved Stellar Populations and their Environments \hspace*{40pt} \linebreak
  $\square$    Galaxy Evolution   \hspace*{45pt} $\square$             Multi-Messenger Astronomy and Astrophysics \hspace*{65pt} \linebreak
  
\textbf{Principal Author:}

Name:	Lynne Valencic
 \linebreak						
Institution:  Johns Hopkins University
 \linebreak
Email: lynne.a.valencic@nasa.gov
 \linebreak
Phone:  301-286-1041
 \linebreak
 
\textbf{Co-authors:} Lia Corrales (University of Michigan); 
Sebastian Heinz (University of Wisconsin); 
Randall K. Smith (SAO);
 Geoffrey C. Clayton (Louisiana State University); 
 Elisa Costantini (SRON); Bruce Draine (Princeton University); Julia Lee (Harvard University); Frits Paerels (Columbia University); Tea Temim (Space Telescope Science Institute); Joern Wilms (University of Erlangen-Nuremberg) \linebreak

\justifying
\textbf{Abstract  (optional):}
The X-ray regime is a largely underused resource for constraining interstellar dust grain models and improving our understanding of the physical processes that dictate how grains evolve over their lifetimes. This is mostly due to current detectors' relatively low sensitivity and high background, limiting the targets to the brightest sources. The improved sensitivity of the next generation of X-ray detectors will allow studies of much fainter sources, at much higher angular resolution, expanding our sampled sightlines in both quality and quantity.

\pagebreak

\section{Introduction}

The effects of dust can be seen along virtually all sightlines -- toward objects in the Milky Way, its neighboring galaxies, and the high-redshift Universe -- and it is necessary to accurately account for them in order to recover an object's intrinsic energy distribution. Dust is also interesting in its
\begin{floatingfigure}[l]{3.4in}
\includegraphics[totalheight=2.1in]{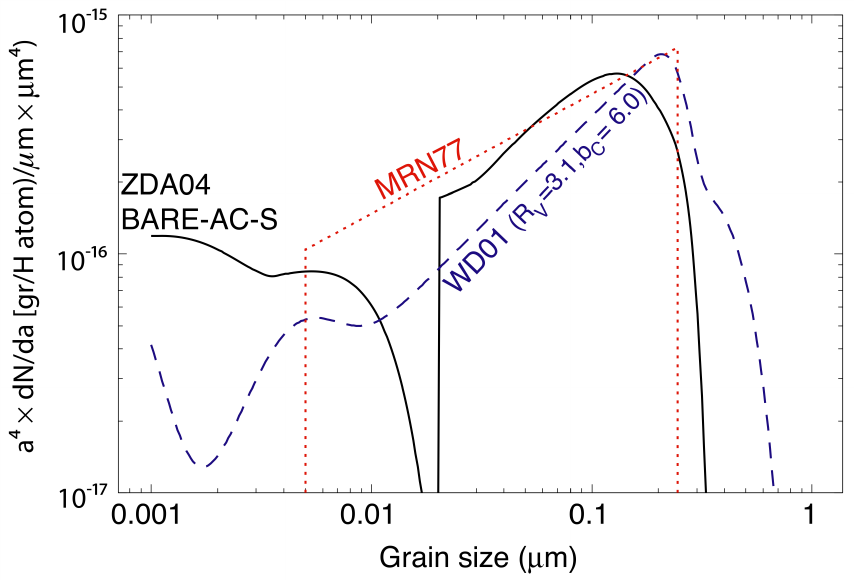}
\caption{\textit{The grain size distributions for three popular models \citep{1977ApJ...217..425M, 2001ApJ...548..296W, 2004ApJS..152..211Z}, weighted 
to show which grain masses dominate. The largest grains show a similar distribution and size range, but produce significantly different scattering. From \citet{2016ApJ...818..143S}.}
\label{fig:sizes_cropped}}
\end{floatingfigure}
\noindent
 own right. Grain surfaces are the sites of molecule growth, and grains themselves are 
 a vital repository for metals in the Universe. 
 They also play an important role in regulating the energy budget of a galaxy, as they generate almost all of the extinction (via both scattering and absorbing) at UV and optical wavelengths, and re-emit this energy in the IR. And ultimately, dust grains are the fundamental building blocks of planet formation.
 
Grains are injected into the ISM from their formation sites in the upper atmospheres of evolved stars and supernova ejecta where they cycle through diffuse and dense environments. In warm/hot media, they are subjected to destructive processes, such as shattering and sputtering. As large grains are preferentially destroyed by shocks,
this leads to a large population of small grains. In cold media, grains grow via coagulation and icy mantle formation. As dense clouds collapse to form stars and planetary systems, the grains are further processed by the proto-stars. Eventually, the stars age to become supernovae or evolved giants, returning their enriched material to the ISM and allowing the ``dust life cycle'' to continue.

Despite the importance of dust, we still do not know some of its most basic properties, such as  grain composition, size, and physical processing in the ISM. In this white paper, we will address how astronomers in the next 10 years will use X-ray imaging of dust scattering halos and echoes to better determine fundamental dust properties, as well as directly measure the distances to bright sources and map the ISM with unprecedented accuracy.

\section{How large are the largest grains, and how common are they?}
\label{sec:halos}
Characterizing grains at the large end of the grain size distribution has important implications for modeling and understanding dust survival in the ISM.
X-ray halos provide a key tool in examining dust grain sizes in the ISM. 
This phenomenon is observed when X-rays from a bright source undergo small-angle scattering off of intervening dust grains in the ISM. The intensity and shape of the halo's radial profile depends on the source brightness and grain characteristics, specifically the composition, size, distribution along the line of sight, and column density \citep{1965ApJ...141..864O, 1991ApJ...376..490M, 2003ApJ...598.1026D}. The scattering cross section increases rapidly with grain size, making halo observations exquisitely sensitive to the large end of the grain size distribution, which itself is dependent on the grain model.

The grain size distributions for some popular models are shown in Fig. \ref{fig:sizes_cropped}. The vast majority of models address the grain characteristics only of those grains in the ``standard diffuse ISM'', with R$_{V}$=3.1, where R$_{V}$ is the ratio of total to selective extinction. Values of R$_{V}$ range from $\sim$2.5 to $\sim$5.5 in the Milky Way, depending on the dominant grain size and thus the grains' environment; low values indicate diffuse media while high values indicate dense clouds \citep{1989ApJ...345..245C}. Thus, only one phase of the grain life cycle is considered by most common models. Further, the models are degenerate, as there are many that can reproduce the various observables associated with dust derived from other wavelength regimes (e.g. Compi\`{e}gne et al. 2011). This highlights the need to 
examine, as much as possible, a broad a section of the spectrum -- from the IR through X-rays -- and sightlines that sample a wide range of environments.

\begin{floatingfigure}[r]{3.8in}
\includegraphics[totalheight=2.4in]{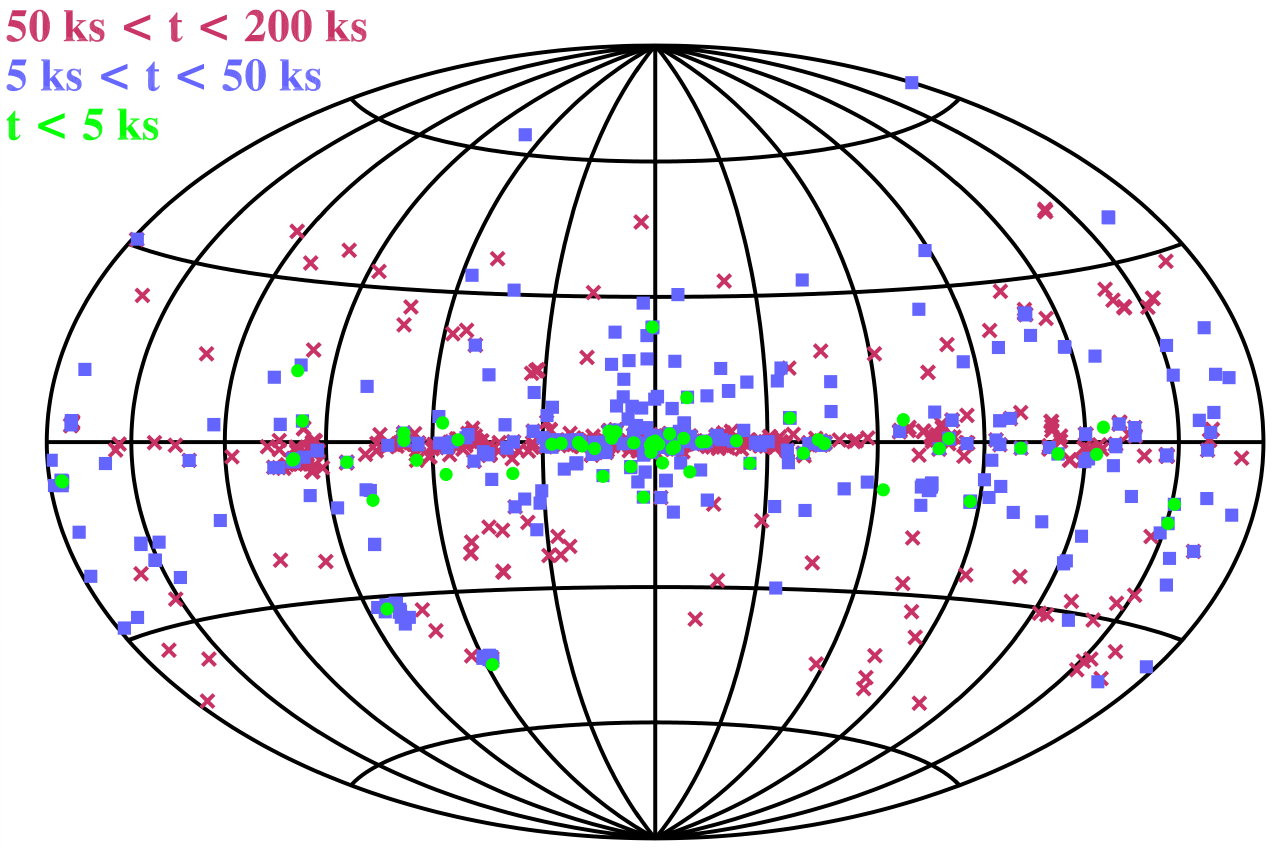}
\caption{\textit{With the next generation of X-ray imagers, dust halos will be easily observable for thousands of sources in reasonable exposure times, over a broad range of environments.}
\label{fig:aitoff_axis}}
\end{floatingfigure}

Many groups have observed halos over the past 40 years \citep[e.g.][]{1983Natur.302...46R, 1995A&A...293..889P, 2015ApJ...809...66V}. But even studies with current detectors are plagued by the problem that is common of all X-ray studies of the ISM: 
\textbf{
the low collecting areas of current X-ray missions prohibits selecting sightlines based on the qualities of the ISM along them or the existence of complementary data at other wavelengths, significantly reducing their utility.}
Even so, they do shine a light on the nature of grains; for instance, porosity has been investigated with conflicting results, and it was found that the diffuse ISM is unlikely to contain a significant presence of grains with radius $a \gtrsim 0.4 \mu$m \citep{1994ApJ...436L...5W, 1995A&A...293..889P, 2015ApJ...809...66V}. This contrasts with the results from  missions to the outer solar system, which found interstellar grains with much larger radii ($a > 1 \mu$m) \citep{1993Natur.362..428G, 2000JGR...10510343L, 2014Sci...345..786W} and raises the possibility of an as-yet mostly undetected population of large grains in the diffuse ISM. 

With their large effective area ($A_E$) and low background, the next generation of detectors will be capable of detecting halos along sightlines chosen for their ISM properties 
and the existence of multiwavelength data, thus allowing truly comprehensive studies of the line-of-sight ISM. Moreover, they will do this in remarkably short exposure times. For instance, it takes about 10$^4$ photons in the halo for solid modeling results (Smith et al. 2002). With Chandra, a moderately bright, absorbed source (F$_X$ = 10$^{-11}$ erg cm$^{-2}$ s$^{-1}$ and N$_{\rm{H}}$ = 5$\times$10$^{21}$ cm$^{-2}$) needs exposure time $t$ = 290 ks to reach the necessary number of halo photons. In contrast, for an imager with $A_E$ = 7000 cm$^{2}$ at 1 keV 
and background = $2\times 10^{-4}$~ct~s$^{-1}$ keV$^{-1}$ arcmin$^{-2}$ at 1~keV, 
it needs only 20 ks. Indeed, halos will be an inevitable by-product of observations. For instance, for an imager with the aforementioned characteristics, $\sim$3700 sources will have measurable halos in $<$ 200 ks in both dense and diffuse media ($10^{20}$ cm$^{-2} < $N$_{\rm{H}} < 5\times10^{22}$ cm$^{-2}$), thus allowing the examination of dust at different phases of its life cycle (Fig. \ref{fig:aitoff_axis}).

In addition to producing halos, small-angle scattering significantly and systematically affects the measured source spectrum, even for lightly absorbed sources. In order to measure this, the grain scattering cross section must be combined with a model that specifies the grain size distribution and composition; 
however, the different size distributions can lead to non-negligible differences in the resulting spectra, particularly in heavily absorbed sightlines \citep{2016ApJ...818..143S}. To illustrate, we modeled the spectrum of the microquasar GRS 1758-258 as a combination of an absorbed disk blackbody and power law \citep{2011MNRAS.415..410S, 2016ApJ...818..143S}. First, we found each component's contribution to the flux without scattering. Then, we included the effects of scattering using different models. The results are in Table \ref{tab:fluxes}. 

\begin{table}[t]
    \centering
    \begin{tabular}{|l|c|c|c|c|}
    \hline
        F$_X$ (10$^{-10}$ erg/cm$^{2}$/s) & No Scattering & (1) & (2) & (3) \\
        \hline \hline
        Disk blackbody & 8.38$\pm$0.10 & 9.04$\pm$0.11 & 9.31$\pm$0.11 & 8.88$\pm$0.10 \\
        Power law      & 1.20$\pm$0.03 & 1.24$\pm$0.03 & 1.23$\pm$0.03 & 1.23$\pm$0.03\\ 
        \hline
    \end{tabular}
    \caption{\textit{A comparison of the contributing fluxes in the spectrum of GRS 1758-258 assuming scattering with different grain models: (1) \citet{1977ApJ...217..425M}, (2) \citet{2001ApJ...548..296W}, (3) the BARE-AC-S model from \citet{2004ApJS..152..211Z}. Fluxes are over the range 0.8-10 keV.}}
    \label{tab:fluxes}
\end{table}

\section{What is the grain destruction rate?}

Studies of halos will also shed light on processes that affect the grain life cycle. A long-standing problem is the uncertainty in the destruction rate, which governs grain lifetimes. Grains are primarily destroyed in supernova (SN) shocks \citep{1980ApJ...239..193D, 1983ApJ...275..652S} and it has been estimated that all grains in the Galaxy's and Magellanic Clouds' ISM should be destroyed in $\sim<$ 10$^{8}$ years \citep{1994ApJ...433..797J, 2014A&A...570A..32B, 2015ApJ...799..158T}. However, the timescale for dust injection into the ISM from stars is on the order of 10$^{9}$ years \citep{1994ApJ...433..797J}. This suggests that there must be a way to construct grains in the diffuse ISM. While it has been shown that carbonaceous grains can re-form, it remains difficult to regrow silicates \citep{2011A&A...530A..44J}, though some laboratory work shows it may be possible \citep{2014ApJ...782...15K}. Another way around this problem is if the dust and gas motions are not coupled, thereby reducing the efficiency of grain destruction in SN shocks \citep{2004ApJ...614..796S}. In this case, grain shattering due to turbulence becomes the dominant means of destruction, and grains may have much longer lifetimes, $\sim$10$^{9}$ years \citep{2016P&SS..133...17H}, in much better agreement with the injection rate. 

Examination of possible halos near neutron stars \citep{2011ApJ...742....4O, 2012ApJ...757...39Y, 2013MNRAS.429.2493C, 2013MNRAS.429.3123E} will shed light on the grain processing in and around SN remnants. Such studies require a telescope with high angular resolution and low background, due to the faintness of the neutron star and halo, and the closeness of the halo to the star. For instance, the neutron star RRAT J1819–1458  
has diffuse X-ray emission extending to 20'' from the star which may be due to dust scattering \citep{2013MNRAS.429.2493C}. In 90 ks, Chandra can detect only 800 photons of diffuse emission. In contrast, a detector with the characteristics described in \S \ref{sec:halos} would collect an order of magnitude more photons in the same time. 

\section{What are the grains' compositions?}

X-ray microcalorimeter technology  enables high resolution spectroscopic imaging at the 2-4 eV
level. This will, for the first time, allow us to measure a high resolution spectrum from a dust scattering halo, which will directly reveal resonant features in the scattering cross-section that can be used to identify different mineral components of interstellar dust grains. Figure~\ref{fig:SiKhalo} shows the
\begin{floatingfigure}[r]{3.7in}
\includegraphics[totalheight=2.3in]{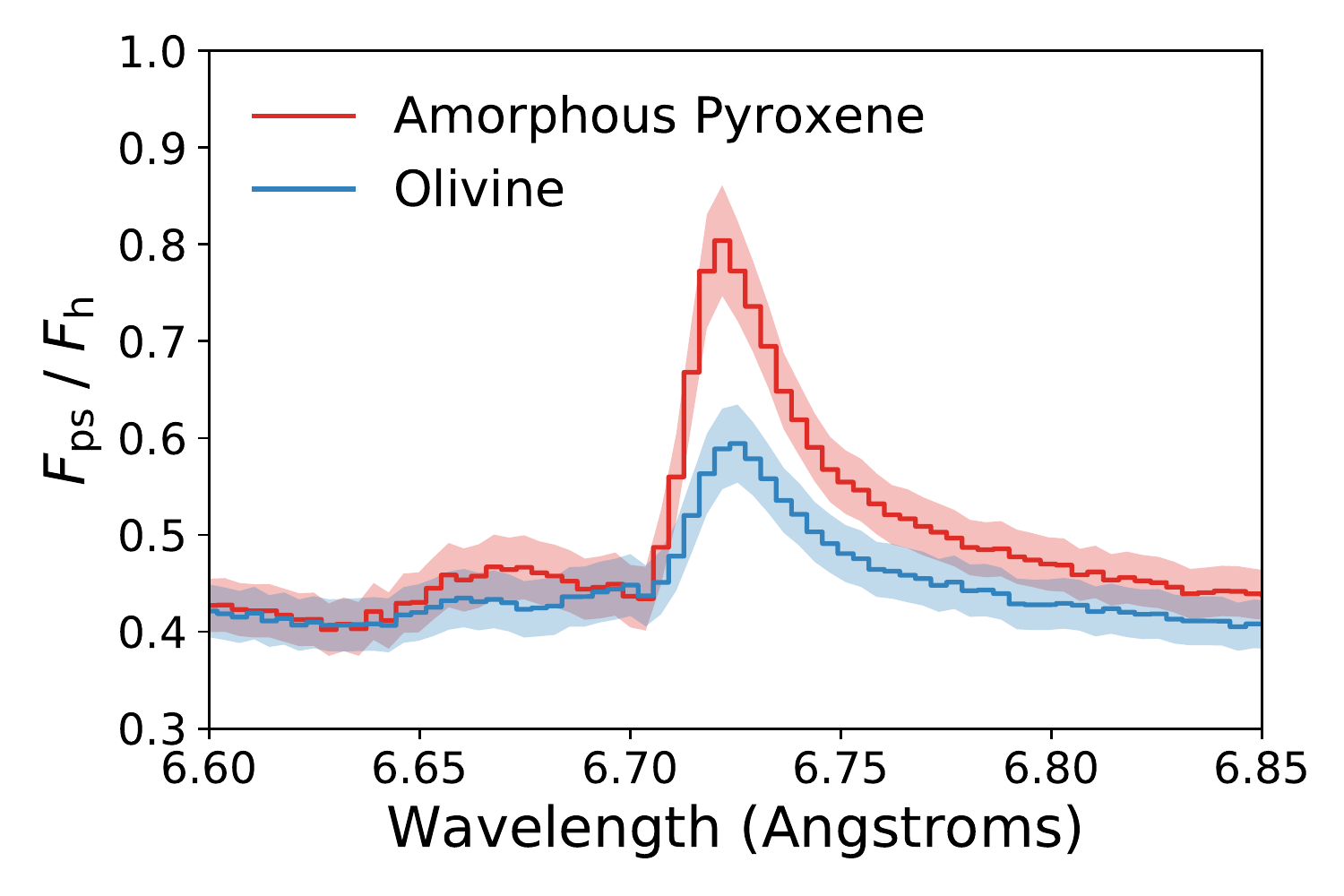}
\caption{\textit{Simulated 50~ks observation of the scattering halo spectrum of GX 340+0 shows a  clear differentiation between grain types.}
\label{fig:SiKhalo}}
\end{floatingfigure}
\noindent
 simulated spectrum of the scattering halo for GX 340+0 ($F_{\rm h}$) normalized by the apparent spectrum of the central point source ($F_{\rm ps}$) using $R \equiv \lambda/\Delta\lambda \sim 1000$, $A_E=10^4$ cm$^{2}$, and the optical constants of \citet{Zeegers2017}. The peak in the scattering cross-section can be used to differentiate between common amorphous and crystalline silicate minerals. Large grains also enhance the height of the scattering peak \citep{Corrales2016,Zeegers2017}. Thus, scattering halo spectra can be used to reveal ices bound up in large dust grains ($\geq 0.3~\mu$m), which, due to shielding, do not exhibit strong absorption \citep{Wilms2000, Jenkins2009}.
The advent of X-ray microcalorimeter instruments with high imaging resolution will also allow us to probe the X-ray scattering spectra from individual ISM clouds, which can be identified through  enhanced or diminished scattering across a $15'$ field of view (FoV)  \citep[e.g.,][]{2015ApJ...806..265H}.

\section{What is the distribution of dust along a line of sight?} \label{sec:distr}
While halos for constant sources are extremely useful diagnostics of the ISM, the halos associated with time-variable sources are perhaps even more so. This is because the scattered X-rays in a halo have a longer path length than non-scattered X-rays, and thus show a time delay. Therefore, if the X-ray source is variable, the halo's radial profile will change, yielding information on the distance to the intervening cloud(s) to extraordinarily high precision, up to $\sim$ 1\% \citep{1973A&A....25..445T, 2015ApJ...806..265H, 2016MNRAS.455.4426V, 2017MNRAS.472.1465P, 2018MNRAS.477.3480J}. 

Time series imaging of such echoes allows the 3-dimensional reconstruction of the column density distribution of dust clouds along the line-of-sight within the limits of the FoV and resolution limits of the X-ray telescope. The unique identification of individual clouds with rings in such X-ray light echoes, together with knowledge of the spectrum of the flare producing the echo, encode the energy and scattering angle dependence of the scattering cross section. With knowledge of the distance to the X-ray source, such echoes can thus be used as sensitive probes of grain sizes and compositions \citep[e.g.][]{2016ApJ...825...15H}. With Gaia's accurate distances across the Galaxy, X-ray tomographic observations of interstellar dust will become precision tools to map out the galaxy in dust and to test the grains' mineralogy and size distributions. Further, it will be possible to use correlations between dust and molecular cloud data to measure streaming motions of gas relative to Galactic rotation, and to constrain the gas-to-dust ratio as a function of Galactic position.

\begin{figure}[t]
\includegraphics[width=\textwidth]{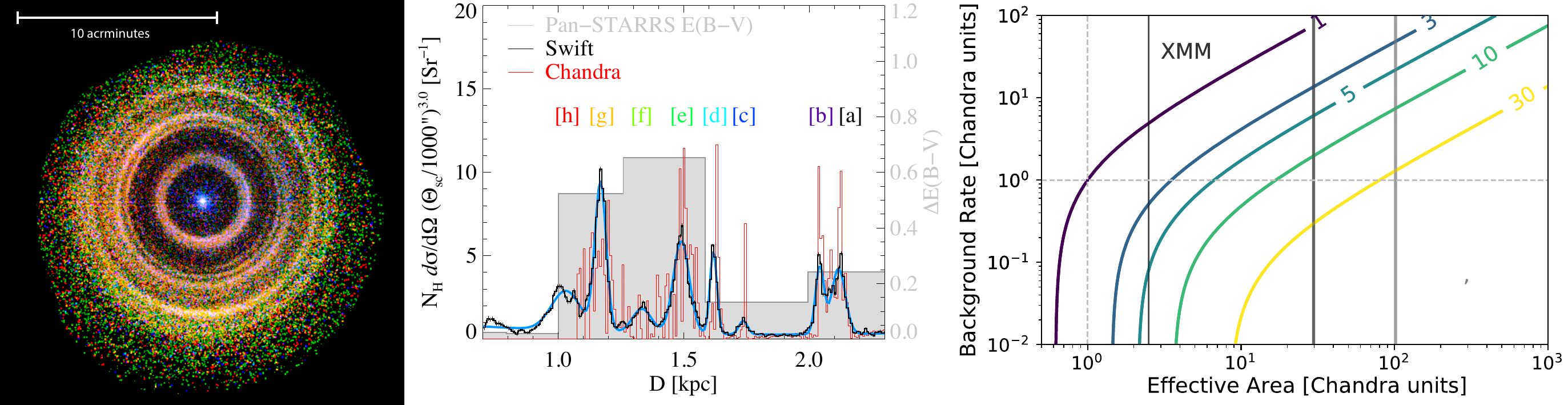}

\caption{\textit{Left: Swift image of the dust echo of V404 Cyg in 2015 \citep{2016ApJ...825...15H}, showing the need for large contiguous FoVs for dust tomography. Middle: Dust column density distribution towards V404 Cyg derived from the echo, showing the vast improvement in precision offered by the high-spatial resolution of Chandra (red) compared in particular to differential extinction measurements (grey). Each letter corresponds to a ring. Right: Dust tomography discovery outlook for increased sensitivity by future missions as a function of both effective area and background rate, compared to Chandra. Contours show the number of echoes detectable compared to Chandra \citep{corrales:19}.\label{fig:tomo}}}
\end{figure}

To enable order-of-magnitude improvements in dust echo tomography, we must plan carefully for the next generation of X-ray telescopes. This includes common requirements for X-ray telescopes: High spatial resolution (comparable to or better than a few arcseconds), 
large FoV, and high A$_E$
in the energy band of 1-5 keV. However, tomography also requires less common capabilities: Executing target of opportunity (ToO) observations with $\sim$1 day turnaround; low diffuse background rates; and the ability to observe potentially bright point sources. Finally, triggering observations of echoes requires all-sky X-ray monitoring, such as currently provided by MAXI.

For a detector with $A_E$ and background as described in \S \ref{sec:halos},
over a lifetime of 5 years, the ISM for $\sim$50 sightlines will be able to be mapped in this way, to greater precision than current stellar population extinction maps; see Fig. \ref{fig:tomo}. 
Furthermore, higher $A_E$ telescopes raise the possibility of observing X-ray scattering from circumgalactic or intergalactic dust \citep{Corrales2015b}.

\section{Conclusion}

X-ray dust scattering halo studies have
the potential to deepen our understanding of grains in diffuse and dense media. Up until now, halo studies have been limited to only the brightest X-ray sources. Modern detectors with large $A_E$, low background, and high angular resolution will allow studies of sightlines that are chosen for the qualities of their ISM and the existence of complementary data at other wavelengths. Further, telescopes with fast ToO capabilities will permit tomography studies, which will produce high-precision dust cloud maps and test grain compositions and size distributions.
This will
lead to more realistic grain models over the range of the grains' life cycle, allowing us to more accurately account for the effects of dust in other astronomical studies, and greater understanding of the physical
processes which dictate the grain life cycle.

\pagebreak
\bibliographystyle{aasjournal}
\bibliography{refs.bib}

\end{document}